\documentclass{fcs1}
\usepackage{bm}
\usepackage{enumerate}
\usepackage{cite}
\usepackage{amsmath}
\usepackage{amscd}
\usepackage{amssymb}
\usepackage{latexsym}
\usepackage{array}
\usepackage{tabularx}
\usepackage[ruled]{algorithm}
\usepackage{algorithmic}
\usepackage{epsfig,subfigure}

\usepackage{multirow}
\usepackage{bigstrut}
\usepackage{array}
\usepackage{epstopdf}
\usepackage{graphicx}

\newcommand{\PreserveBackslash}[1]{\let\temp=\\#1\let\\=\temp}
\newcolumntype{C}[1]{>{\PreserveBackslash\centering}p{#1}}
\newcolumntype{R}[1]{>{\PreserveBackslash\raggedleft}p{#1}}
\newcolumntype{L}[1]{>{\PreserveBackslash\raggedright}p{#1}}

\newtheorem{ther}{Theorem}
\newtheorem{deft}{Definition}

\volumn{ }
\doi{ }
\articletype{REVIEW~ARTICLE}
\copynote{{\copyright} Higher Education Press and Springer-Verlag Berlin Heidelberg 201X}
\ratime{Received month dd, yyyy; accepted month dd, yyyy}
\email{$scong@ustc.edu.cn$}
\title{
State of the Art and Prospects of Structured Sensing Matrices in Compressed Sensing}
\author{Kezhi Li $^{1}$, Shuang Cong \xff $^{2,*}$}
\address{{$^1$ACCESS Center, Royal Institute of Technology (KTH) \\     Stockholm, 10044, Sweden}\\
{$^2$Dept. of Automation, University of Science and Technology of China  \\ Hefei, 230027, China}}

\begin{document}
\maketitle
\setcounter{page}{1}
\setlength{\baselineskip}{14pt}

\begin{abstract}
Compressed sensing (CS) enables people to acquire the compressed measurements directly and recover sparse or compressible signals faithfully even when the sampling rate is much lower than the Nyquist rate. However, the pure random sensing matrices usually require huge memory for storage and high computational cost for signal reconstruction. Many structured sensing matrices have been proposed recently to simplify the sensing scheme and the hardware implementation in practice. Based on the restricted isometry property and coherence, couples of existing structured sensing matrices are reviewed in this paper, which have special structures, high recovery performance, and many advantages such as the simple construction, fast calculation and easy hardware implementation. The number of measurements and the universality of different structure matrices are compared.
\end{abstract}


\section{Introduction}

In the digital revolution, people are now employing various signal processing techniques and new sensing systems in general electronic products with ever-increasing resolution and fidelity. The conventional manners of sampling signals, images, videos, or other data obey the celebrated Shannon's theorem, that requires to sample a signal at a sampling rate at least twice the highest frequency present in a signal (so-called Nyquist rate) to retain signal information intact \cite{Shannon-Comm,Nyquist-Certain}. The Shannon's theorem solves the problem in theory perfectly, yet unfortunately it is not omnipotent. In many applications such as remote surveillance or spectroscopy, sampling in the result with Nyquist rate is expensive, or even physically impossible. So as long as the recovery performance achieves an acceptable level, people want to build devices which are capable of acquiring samples at a necessary rate as low as possible. In some other applications, such as imaging system or video processing, sampling a large number of measurements seems feasible. However, because of the limited storage space and using of advanced compression techniques, people often discard the most received data, and just save a small amount of the compressed data (e.g., JEPG). Apparently it will waste lots of valuable sensing resources since the entire data are sampled at first.

Aiming at solving above problems, the compressed sensing (short for CS) theory \cite{donoho-cs,robust-uncertain,Universal-decoding,Candes-AnIntroToCS,Baraniuk-CS} has become one of the hottest research areas in signal processing since 2006. The research of CS has been growing very fast and it focuses on acquiring and reconstructing sparse or compressible signals. By using CS, compressed measurements can be acquired directly and one may recover the original sparse or compressible signal faithfully even when the sampling rate is much lower than the Nyquist rate.
An $N$-length signal $\mathbf{x}$ is regarded as sparse if $\mathbf{x}$ has $K$ nonzero values and $K  \ll N$. Compressible $\mathbf{x}$ means that $\mathbf{x}$ can be well-approximated by another sparse signal $\mathbf{f}$ in certain domain $\mathbf{\Psi}$ by using only $K$ nonzero coefficients: ${\mathbf{x}}={\bf \Psi} \mathbf{f}, |\mathbf{f}|_0 = K$. Normally the traditional compression techniques preserve the values and locations of the largest coefficients, such as JPEG, JPEG2000, MPEG. While CS has more efficient sensing or sampling protocols that capture the essential information content embedded in the original signal and obtain the condensed data straightforwardly. More precisely, these protocols are nonadaptive linear transforms, which can be represented by well-designed matrices, called \emph{sensing matrices} $\mathbf{\Phi}$. These matrices should be incoherent to the sparsifying matrix $\mathbf{\Psi}$ of the compressible signal. With the measurements and the sensing matrix, the process of exact reconstructing signals from a subset of measurements can be implemented by solving a nonlinear optimization problem. The approaches to solving the nonlinear problems are named \emph{reconstruction/recovery algorithms}. From a mathematical point of view, compressed sensing is also deemed as a technique of finding sparse solutions to underdetermined linear systems.

The CS theory is a revolution in both the theory of reliable signal sampling and physical design of sensors. Since the original signal can be sensed from fewer linear projections rather than acquired in its initial domain, the \emph{sensing matrices} play an important role in the CS framework. The property of the sensing matrices affects the number of necessary measurements and the recovery performance directly. Early researchers proved that a random projection is one of the best solutions \cite{Universal-decoding, prac-cs}. The projection matrices are generated by orthogonalizing measured vectors uniformly and independently on a unit sphere. In addition, sensing matrices consisting of independent and identically distributed (i.i.d) entries drawn from a Gaussian or Bernoulli distribution also perform well in both theory and practice \cite{donoho-cs,robust-uncertain}. Though the problem has been solved in mathematics, there still exist many obstacles to overcome. One main drawback of the pure random sensing matrices is that they require huge storage-memory, namely $M \times N$ entries to recover a length-$N$ signal, and high computational cost for signal reconstruction. Moreover, the difficulty of hardware implementation also makes them expensive in practice.

To simplify the sensing scheme, many structured sensing matrices have been proposed in recent years. In this paper, after explaining some terminologies such as restricted isometry property and coherence, we give an introduction to couples of existing structured sensing matrices, including subsampled incoherent bases, random Toeplitz matrices, random demodulator matrices, random convolution matrices, structurally random matrices, convolutional matrices using sequences and some other structured sensing matrices. These matrices have special structures which equip them efficiency in the construction, calculation or hardware implementation. For many of them, corresponding fast recovery algorithms have been developed by exploiting their specific structures. Here we put our emphasis on how the structured matrices are generated and what are their recovery performances. Based on these, the number of measurements and the universality of different structure matrices are compared. The paper will help readers to understand the characteristics of popular sensing matrices well and may inspire them to explore or pursuit more efficient sensing schemes in the CS area.

The remaining paper is organized as follows. In Section 2, we describe the core concepts of this paper: sensing matrices, and introduce two prevalent criteria that examine the effectiveness of sensing matrices: restricted isometry property and coherence. In Section 3,  couples of structured sensing matrices are analyzed. The overview of applications is discussed in Section 4. Finally, the prospects of structured sensing matrices are discussed in Section 5 followed by the conclusions in Section 6.

\section{Sensing Matrices}


In our real world, normally the useful signals are not random. Images, videos or voices often contain specific structures and strong correlation among pixels, frames or samples. These structures and correlations are the assumptions behind the sparse representation theory. Given an $N$-dimensional signal $\mathbf{x}$, $\mathbf{\Psi}$ denotes the sparsifying transform basis for $\mathbf{x}$, where throughout this paper we assume that ${\bf \Psi}$ is an $N\times N$ normalized unitary
matrix satisfying ${\bf \Psi}^{*}{\bf \Psi}=N{\mathbf{I}_N}$. So $\mathbf{x}$ can easily be decomposed by means of a linear superposition of $K$ elementary components:
\begin{equation}\label{eq:x=PsiF1}
    {\mathbf{x}}=\sum_{k=1}^{K}f_k \psi_k.
\end{equation}
which can be rewritten in the form of matrix multiplication as
\begin{equation}\label{eq:x=PsiF2}
    {\mathbf{x}}={\bf \Psi} \mathbf{f},
\end{equation}
where $\mathbf{f}$ is a length $N$ sparse vertical vector with $K$ nonzero values, $K \ll N$. Typical transforms $\mathbf{\Psi}$ include discrete Fourier transform (DFT), discrete cosine transform (DCT) and Discrete Wavelet Transform (DWT). Sometimes the number of nonzero values in $\mathbf{f}$ is larger than $K$. In this case people usually encode the most significant $K$ non-zero entries of $\mathbf{f}$ and disregard the rest, which is also the core principle of the image compression standard JPEG (using DCT) and JPEG2000 (using DWT).

\begin{figure*}[tp]
\mbox{}\\
   \centering
   \includegraphics[width=12.8cm]{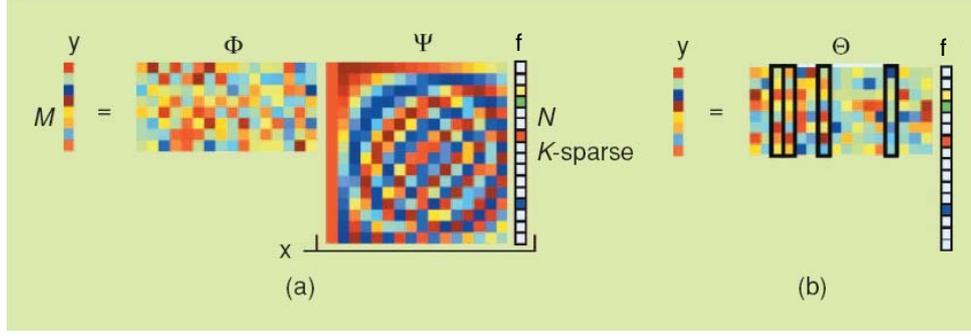} 
  \caption[(a) Compressive sensing measurement process with a random Gaussian
measurement matrix $\mathbf{\Phi}$ and DCT matrix $\mathbf{\Psi}$ as sparsifying matrix. (b) Measurement process with $\mathbf{\Theta} = \mathbf{\Phi \Psi}$. ]{(a) Compressive sensing measurement process with a random Gaussian
measurement matrix $\mathbf{\Phi}$ and DCT matrix $\mathbf{\Psi}$ as sparsifying matrix. (b) Measurement process with $\mathbf{\Theta} = \mathbf{\Phi \Psi}$. The original scheme figure is from \cite{Baraniuk-CS}. We use $\mathbf{f}$ to denote the $K$ sparse vector.
 }\label{fig:CS-diagram}
\end{figure*}

Now assuming a length-$N$
signal ${\mathbf{x}}$ as defined in (\ref{eq:x=PsiF1}), the
data acquisition process can be described as
\begin{equation}\label{eq:CS-measure-basic1}
\mathbf{y}={\bf \Phi} \mathbf{x}={\bf \Phi} {\bf
\Psi}\mathbf{f} = \mathbf{\Theta} \mathbf{f},
\end{equation}
where the measurement $\textbf{y}$ represents an $M \times 1$ sampled vector, ${\bf
\Phi}$ is an $M\times N$ measurement/sensing matrix, $\mathbf{\Theta}= {\bf \Phi} {\bf
\Psi}$. (\ref{eq:CS-measure-basic1}) is the kernel equation of the sensing process. The sensing process with a random Gaussian measurement matrix $\mathbf{\Phi}$ and a DCT matrix $\mathbf{\Psi}$ is illustrated in Figure \ref{fig:CS-diagram}, in which there are
four columns of $\mathbf{\Psi}$ that correspond to nonzero $\mathbf{f}_i$ coefficients; the measurement vector $\mathbf{y}$ is a linear combination of these columns. The CS theory considers problems based on the fundamental equation (\ref{eq:CS-measure-basic1}). These problems can be summarized as how to design efficient sensing matrix $\mathbf{\Phi}$, and how to recover $\mathbf{x}$ given $\mathbf{y}$ and $\mathbf{\Phi}$.

Here we focus on the problem of designing proper sensing matrices. Conventionally, if $K$ entries in $\mathbf{y}$ are more important than other entries, people may capture the signal energy roughly and recover the original signal from $K$ measurements, like what we do for recovering natural image from its frequency spectrum.
However, in the CS framework the entries of $\mathbf{x}$ are assumed sparse and randomly distributed,
which means people do not know where the large entries locate. In this circumstance $\mathbf{x}$ is able to be recovered by exploiting the sparsity from $M$ measurements, $M \ll N$ and $M>K$. Because only partial $M$ measurements are captured to recover the original signal, a good sensing scheme should spread out the information of the non-zero entries into every measurement $\mathbf{y}_k$ evenly, in case of losing significant information. Follow this intuitive idea, people found that random projection is one of the best candidates as a sensing matrix \cite{Universal-decoding,Haupt-SignalRecons}. Besides, if ${{\bf \Phi}}$ represents a Gaussian or Bernoulli random operator,
${\mathbf{x}}$ can also be faithfully recovered from ${\mathbf{y}}$ using
nonlinear optimization approaches provided that $M\geq
\mathcal{O}(K\log(N/K))$ \cite{donoho-cs,robust-uncertain,Cand-Quantitative}. These early works made by D. Donoho, E. Cand\`{e}s, T. Tao and Romberg established the foundation of the CS theory.

From Eq. (\ref{eq:CS-measure-basic1}) an essential question might be raised instinctively: apart from the general random operators, what kinds of  sensing matrices $\mathbf{\Phi}$ are capable to recover $\mathbf{x}$ uniquely from measurements $\mathbf{y}$? Fortunately, two important criteria for evaluating proper operators were created to provide fundamental insights into the geometry of sensing matrices. The most well-known one is often referred as the Restricted Isometry Property (RIP):

\begin{deft}[RIP\cite{Universal-decoding}]\label{deft:RIP}
An $M\times N$ matrix ${\bf \Theta}={{\bf \Phi}}{\mathbf{{\bf
\Psi}}}$ is said to satisfy the RIP with parameters $(K, \delta)$
($\delta \in (0,1)$) if
\begin{equation}\label{eq:RIP}
(1-\delta)\|{\mathbf{f}}\|^2\leq \|{\bf \Theta}{\mathbf{f}}\|^2 \leq
(1+\delta)\|{\mathbf{f}}\|^2, \text{ for all } {\mathbf{f}} \in
\Gamma,
\end{equation}
where $\Gamma$ represents the set of all length-$N$ vectors with $K$ non-zero coefficients.
\end{deft}
 Generally speaking, RIP requires the sensing matrix to act as a near isometry on the set of all $K$-sparse signals. It is consistent with the thought of spreading energy behind random sensing matrices. So measurement $\mathbf{y}$ preserves the energy that does not shrink or expand too much comparing with the original signal $\mathbf{x}$. If ${\bf \Theta}$ satisfies the RIP, many reconstruction algorithms can be used to recover any $K$-sparse signal $\mathbf{f}$ from $M$ measurements $\mathbf{\Theta f}$, such as Basis Pursuit (BP) or Matching Pursuit recovery algorithms \cite{Donoho-StableRec,Tropp-OMP}. In addition, RIP guarantees the uniqueness of the reconstruction result $\mathbf{f}$, which does not hold automatically for some other RIP-related property, such as the weaker Statistical Restricted Isometry Property (StRIP) \cite{Large-class-STRIP}.

Because there is no existing algorithm for efficiently verifying whether a matrix satisfies RIP, people also need the coherence property to examine the ``quality" of $\mathbf{\Theta}$:

\begin{deft}[Coherence\cite{Donoho-OptimallySparse, Tropp-GreedIsGood}]
The coherence $\mu(\mathbf{\Theta})$ is the largest absolute inner product between any two normalized columns of $\mathbf{\Theta}$
\begin{equation}\label{eq:Coherence1}
\mu(\mathbf{\Theta}) = \max_{1 \leq i \neq j \leq N} {|\langle \Theta_i , \Theta_j \rangle} |
\end{equation}
where $\Theta_i , \Theta_j$ represent two columns of $\mathbf{\Theta}$.
\end{deft}
If $\mathbf{\Theta} = \mathbf{\Phi} \mathbf{\Psi}$, the coherence can also be quantified by calculating the maximal correlation among all rows of $ \mathbf{\Phi}$ and all columns of $\mathbf{\Psi}$
\begin{equation}\label{eq:Coherence2}
\mu(\mathbf{\Phi}, \mathbf{\Psi}) = \max_{1 \leq i, j \leq N} {|\langle\mathbf{\Phi}_i, \mathbf{\Psi}_j \rangle} | = \max_{1 \leq i, j \leq N} {| \mathbf{\Theta} (i,j)} |.
\end{equation}
Note that for a unitary matrix $\mathbf{\Phi}$ with
$\mathbf{\Phi}^{*}\mathbf{\Phi}=N{\mathbf{I}_N}$, the mutual coherence coefficient $\mu$ is bounded by $1\leq \mu(\mathbf{\Theta}) \leq \sqrt{N}$ \cite{Candes-AnIntroToCS}. When $\mathbf{\Phi}$ is
chosen as the DFT or the Walsh-Hadamard transform and $\mathbf{\Psi}$ is an identity matrix, $\mu(\mathbf{\Theta})=1$. If $\mathbf{\Phi}$ is a matrix of random basis vectors or a matrix of i.i.d. Gaussian entries $\mathcal{N}(0,1)$, the mutual coherence between $\mathbf{\Phi}$ and any orthonormal matrix $\mathbf{\Psi}$ is on the order of $\mathcal{O}\left( \sqrt{2 \log N} \right)$ with very high probability, far from the lower bound \cite{Candes-Near-ideal}. Coherence $\mu$ is a core concept in constructing CS matrix, and it will be frequently used in the sensing matrix analysis.

\section{Structured Sensing Matrices}

The initial work of CS focused on randomized sensing matrices, in which the entries of matrices are independently generated from standard probability distributions. For instance, with overwhelming probability, all matrices satisfying random Gaussian/Bernoulli distribution obey the RIP could be uniquely recovered from number of measurements $M$ and
\begin{equation}
M \geq  C \cdot K\log(N/K)
\end{equation}
where $C$ is some constant depending on each instance \cite{robust-uncertain}. As mentioned in the introduction, pure random matrices are not easily applicable to real implementations due to its large storage and heavy computation. In recent years some structured sensing matrices have been proposed. Unlike pure random matrices, special constructions make structured sensing matrices suitable for various applications, and we will introduce them chronologically and analyze their performances respectively.

\subsection{Subsampled Incoherent Bases}

For subsampled incoherent base matrices, the most famous examples are random subsampled Fourier and Walsh-Hadamard matrices. An $M \times N$ sensing matrix is constructed by random selecting rows from an $N \times N$ square DFT (or FFT) matrix $\mathbf{F}$ or a Walsh-Hadamard transform (WHT) matrix $\mathbf{H}$, respectively. Specifically,
\begin{equation}
    \mathbf{F} = \frac{1}{\sqrt{N}} \left[
    \begin{array}{ccccc}
    1 & 1 & 1 & \cdots & 1 \\
    1 & \omega & \omega^2 & \cdots & \omega^{N-1} \\
    1 & \omega^2 & \omega^4 & \cdots & \omega^{2(N-1)} \\
    \vdots & \vdots & \vdots & \ddots & \vdots \\
    1 & \omega^{N-1} & \omega^{2(N-1)} & \cdots & \omega^{(N-1)(N-1)}
      \end{array}
    \right],
\end{equation}
where $\omega= e^{-\frac{2\pi i}{N}}$  is a primitive $N$th root of unity in which $i = \sqrt{-1}$ and
\begin{equation}\label{eq:H_N}
    \mathbf{H}_N=  \left[ \begin{array}{cc}
\mathbf{H}_{N/2} & \mathbf{H}_{N/2}  \\
\mathbf{H}_{N/2} & -\mathbf{H}_{N/2}
\end{array}\right],
\end{equation}
with an initial matrix $\mathbf{H}_1= \left[ +1 \right]$. The oversampling factor for partial DFT matrix was proved as $(\log N)^6$ at first in \cite{Universal-decoding}, then was improved to $(\log N)^4$ in \cite{Rudelson-OnSparse}. Generally speaking, the RIP property of sampled unitary matrix is summarized as following
theorems.


\begin{ther}[RIP for randomly subsampled unitary matrix \cite{Rudelson-OnSparse, Duarte-Structured}]\label{ther:RIP_unitary}
Suppose that the $M\times N$ matrix ${\bf \Theta}$ is a randomly
subsampled unitary matrix, i.e., it can be written as ${\bf
\Theta}=\frac{1}{\sqrt{M}}{\mathbf{R}}_{\Omega}{\mathbf{U}}$, where
 $\frac{1}{\sqrt{M}}$ is a normalizing coefficient,
${\mathbf{R}}_{\Omega}$ is a random
sampling operator which selects $M$ samples out of $N$ ones uniformly
at random, and $\mathbf{U}$ is an $N\times N$ unitary matrix
satisfying $\mathbf{U}^{*}\mathbf{U}=N\mathbf{I}_N$. Then ${\bf
\Theta}$ satisfies the RIP with high probability provided that

\begin{equation}\label{eq:RIPpartialbound}
M\geq \mathcal{O}\left(\delta^{-2}\mu^2(\mathbf{U})K\log^4 N\right).
\end{equation}
\end{ther}
where $\delta$ denotes the restricted isometry constant in definition \ref{deft:RIP}.

Theorem \ref{ther:RIP_unitary} implies that the RIP bound of a randomly
subsampled unitary matrix depends on $\mu({\mathbf{U}})$. Note that
for a unitary matrix ${\mathbf{U}}$ with
${\mathbf{U}}^{*}{\mathbf{U}}=N{\mathbf{I}_N}$, $1\leq
\mu(\mathbf{U}) \leq \sqrt{N}$. When ${\mathbf{U}}$ is
chosen as the FFT or the Walsh-Hadamard transform,
$\mu(\mathbf{U})=1$ and by Eq.~(\ref{eq:RIPpartialbound}), one has
\begin{equation}\label{eq:RIPpartialFFT}
M\geq \mathcal{O}\left(\delta^{-2}K\log^4 N\right).
\end{equation}

All above bounds are for the uniform reconstruction, which means that once the sampling operator ${{\bf
\Phi}}$ is constructed, all sparse signals in a certain basis ${\bf
\Psi}$ can be recovered as long as $M$ is sufficiently large. If one fixes $\mathbf{x}$
and wants to recover it specifically, the problem turns to a non-uniform one and
this weaker assumption leads to less measurements. In detail:
\begin{ther}[Non-uniform recovery \cite{candes-ripless}]\label{ther:non-uniform}
Assume that ${\bf \Theta}$ is a randomly subsampled unitary matrix
that follows the same definition as in
Theorem~\ref{ther:RIP_unitary}. Let ${\mathbf{f}}$ in
(\ref{eq:CS-measure-basic1}) be a fixed arbitrary $K$-sparse signal. Then
${\mathbf{f}}$ can be faithfully recovered from ${\mathbf{y}}$ using
$l_1$ norm optimization, if $M$ satisfies

\begin{equation}\label{eq:non-uniform}
M \geq \mathcal{O}(\mu^2({\mathbf{U}})K\log N).
\end{equation}
\end{ther}

In addition, if we fix $\mathbf{f} \in \mathbf{R}^N$
and suppose that the coefficient sequence $\mathbf{f}$ of $\mathbf{x}$ is $K$-sparse in
the basis $\mathbf{\Psi}$; select $M$ measurements in the $\mathbf{\Phi}$
domain uniformly at random, then if
\begin{equation}
M \geq C \cdot \mu^2(\mathbf{\Phi} , \mathbf{\Psi}) K\log N
\end{equation}
for some positive constant $C$, its $l_1$ norm minimization solution is exact with
overwhelming probability \cite{Candes_sparsity_incoherence, Candes-AnIntroToCS}. For
the cases of DFT and WHT matrices, $\mathbf{\Phi}= \mathbf{F}$ or $\mathbf{H}$, the bound of $M$ holds for $\mathcal{O}(K \log N)$. The theorems listed
here are also very useful to prove the feasibility of other structured sensing matrices.

Although partial FFT (or WHT) has near-optimal theoretical guarantee, easy hardware
implementation and fast-computable recovery, its major shortcoming
is the lack of the \emph{universality} property. A universal sensing matrix means
that the matrix can handle signals that are sparse in any domain. If ${\bf
\Phi}$ is a Gaussian random matrix, the matrix ${\bf \Phi}{\bf
\Psi}$ will remain Gaussian for any unitary transform ${\bf \Psi}$.
However, if ${\bf \Phi}$ is randomly sampled from a FFT, it will not
be universal, as $\mu({\mathbf{F}{\bf \Psi}})$ can not be
$\mathcal{O}(1)$ for all bases ${\bf \Psi}$, eg. when ${\bf \Psi} = \mathbf{F}^*$, $\mu({\mathbf{F}{\bf \Psi}})$ will be large.

\subsection{Random Toeplitz Matrices}

Because all elements in random matrices are required to satisfy the i.i.d. random distribution, it becomes natural to raise a question one step further: can we reduce the randomness a little and achieve a similar reconstruction performance? Bajwa et. al. first followed this thought to propose random Toeplitz matrices (RTM) in 2007 \cite{Toeplitz-CS, Haupt-Toeplitz}. In RTM, the entries are independence distributed in one row, while reserve certain structure among other rows. Specifically, if a probability distribution $P(a)$ yields
an i.i.d. CS matrix (having unit-norm columns in expectation) then an
$M \times N$ (partial) Toeplitz matrix $\mathbf{A}$ (also having unit-norm columns in
expectation) of the form

\begin{equation}
\mathbf{A}= { \left[ \begin{array}{ccccc}
a_{N-1} & a_{N-2} & \cdots & a_{0}  \\
a_{N} & a_{N-1} & \cdots & a_{1} \\
\vdots & \vdots & \ddots & \vdots \\
a_{N+M-2} & a_{N+M-3} & \cdots & a_{M-1}
 \end{array} \right]},
\end{equation}
where the entries $\{a_i \}_{i=1}^{N+M-2}$ have been drawn independently from
$P(a)$, is also a CS matrix in the sense that it satisfies RIP of order $3K$
with high probability for every $\delta \in (0, 1/3)$ provided $M > C
\cdot K^3 \log (N/K)$, where $C$ is a constant \cite{Toeplitz-CS}.

In the technical aspect, the proof of RIP of RTM used the celebrated Hajnal-Szemeredi theorem on equitable coloring of graphs to partition an $M \times 3K$ Toeplitz-structured submatrix $\mathbf{A}_T$ into roughly $\mathcal{O} (K^2)$ i.i.d. submatrices having dimensions approximately equal to $\mathcal{O}(M/K^2) \times 3K$. By using random Toeplitz matrices, only $\mathcal{O}(N)$ independent random variables are required to generate.  Multiplication with Toeplitz matrices can be more efficiently implemented using
fast Fourier transform, resulting in faster acquisition and reconstruction algorithms. In addition, Toeplitz-structured matrices meet the naturally requirement for certain application areas such as system identification. Later Haupt et. al. and Rauhut improved the bound of $M$ to $\mathcal{O}(K^2 \log N)$ \cite{Haupt-Toeplitz} and $\mathcal{O}(K \log^2(N))$ \cite{Rauhut-CirculantAnd}, respectively.

Meanwhile, random Toeplitz matrices also have disadvantages. For example, RTM are proved to be able to recover signals sparse only in the time domain. Their strong structures make them not suitable for processing signals sparse in other bases, such as DCT domain.

\subsection{Random Demodulator}

\begin{figure*}[tp]
\mbox{}\\
   \centering
   \includegraphics[width=10cm]{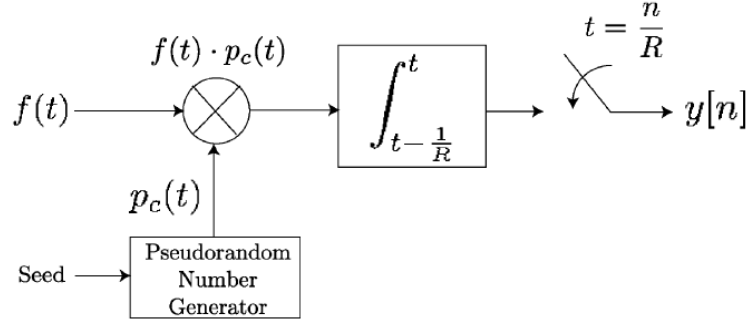} 
  \caption[Block diagram for the random demodulator. The components include a
random number generator, a mixer, an accumulator, and a sampler.]{Block diagram for the random demodulator. The components include a
random number generator, a mixer, an accumulator, and a sampler (taken from \cite{Tropp-BeyondNyquist}).
 }\label{fig:RD-diagram}
\end{figure*}

The random demodulation (RD) matrix was proposed by Tropp et. al. in 2010 \cite{Tropp-BeyondNyquist}. Pseudorandom binary sequence are often used to modulate the input signal. Similar implementations include Bernoulli or Rademacher random variables. The random demodulator is a sampling system that can be used to acquire sparse, bandlimited signals in an analog model. Fig. \ref{fig:RD-diagram} displays a block diagram for the RD system \cite{Tropp-BeyondNyquist}. It is for a continuous-time signal $f$ whose highest frequency is less than $W/2$ hertz. Tropp et. al. modulated the signal by multiplying the signal with a high-rate pseudonoise sequence, which smeared the tones across the entire spectrum. Then a low-pass anti-aliasing filter was applied to capture the signal $\mathbf{x}$ by sampling $\mathbf{x}$ at a relatively low rate. Simulations suggested that the RD requires just $\mathcal{O} (K \log(W/K))$ samples per second to stably reconstruct the original signal.

In mathematics, the random demodulator can be seen as a linear system that maps a continuous-time signal to a discrete sequence of samples. To express the system in matrix form, let $\varepsilon_0, \varepsilon_1, \cdots, \varepsilon_{W-1} $ be the chipping sequence in a diagonal matrix $\mathbf{D}$, $\mathbf{H}$ is an $R \times W$ accumulate-and-dump sampler matrix, where $R$ is the sampling rate. Assume that $W$ is divisible by $R$, the overall action of the system is
\begin{equation}
\begin{split}
\mathbf{\Theta} = \mathbf{\Phi \Psi} = \mathbf{HD} \cdot \hat{\mathbf{F}},
\end{split}
\end{equation}
where
\small
\begin{equation}
 \mathbf{H}={ \left[ \begin{array}{ccccccccccccc}
1 & 1 & \cdots  &  &  & & & & & &  &  \\
 &  & &  & 1 & 1 & \cdots &   &  & & &  \\
 &  &  &  & & & &\ddots & \ddots & \ddots & & & \\
 &  &  & & & & & &  & 1 & 1 & \cdots
 \end{array} \right]},
 \end{equation}
 there are $W/R$ $1$ in each row of $\mathbf{H}$, and
 \begin{equation}
 \mathbf{D} =  { \left[ \begin{array}{cccc}
\varepsilon_0 &  &  &   \\
 & \varepsilon_1 &  &  \\
 &  & \ddots &  \\
 &  &  & \varepsilon_{W-1}
 \end{array} \right]},
 \end{equation}
 \normalsize
and $\hat{\mathbf{F}}$ is a $W \times W$ permuted DFT matrix with
\small
\begin{equation}
\hat{\mathbf{F}} = \frac{1}{\sqrt{W}} \left[ e^{-2 \pi i \cdot n w/W}\right]_{n,w},
\end{equation}
\normalsize
where $n=0,1, \cdots, W-1$ and $w = 0, \pm 1, \cdots, \pm (W/2-1), W/2$.

The main advantage of the RD system is it bypasses the need for a high-rate analog-to-digital converter (ADC). It is typically much easier to implement demodulation rather than sampling, thus a low-rate ADC is allowed to use and a more robust system with low-power can be achieved. In theory, the RD guaranteed the recovery of random general signals with the sampling rate of $R \sim \mathcal{O} \left(K\log{W} + K \log^3{W} \right) $ in the noiseless case and $R \sim \mathcal{O} ( K \log^6{W})$ in the noisy case, where $C$ is a positive constant.

\subsection{Random Convolution}

\begin{figure*}[tp]
   \centering
   \includegraphics[width=10cm]{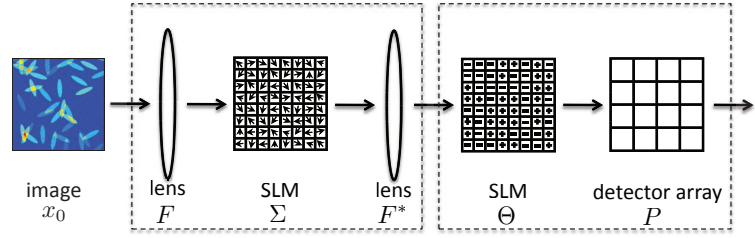} 
  \caption{Fourier optics imaging architecture implementing random convolution followed by RPMS. \cite{random-convolution}. SLM represents the spatial light modulator.
 }\label{fig:RC-diagram}
\end{figure*}

The random convolution (RC) model was first proposed by Romberg in 2007 \cite{Romberg-SensingBy, random-convolution}. In the RC, the construction has two steps. The signal $\mathbf{x} \in \mathbb{R}^N$ was circularly convolved with a ``pulse'' $\mathbf{h} \in \mathbb{R}^N$, then subsampled. The pulse is supposed to be random and its energy spreads uniformly across the discrete spectrum. If one writes the convolution of $\mathbf{x}$ and $\mathbf{h}$ into the matrix form as $\mathbf{H x}$, where \cite{random-convolution}
\begin{equation}
 \mathbf{H} = N^{-1/2} \mathbf{F}^* \mathbf{\Sigma} \mathbf{F},
\end{equation}
with $\mathbf{F}$ as the discrete Fourier matrix and $\mathbf{\Sigma}$ as a diagonal matrix whose non-zero elements are the Fourier transform of $\mathbf{h}$. The matrix $\mathbf{\Sigma}$ can be generated by
\begin{equation}
\mathbf{\Sigma} =   { \left[ \begin{array}{cccc}
\sigma_0 &  &  &   \\
 & \sigma_1 &  &  \\
 &  & \ddots &  \\
 &  &  & \sigma_{N-1}
 \end{array} \right]},
\end{equation}
where the diagonal entries $\sigma_w$ are unit magnitude complex numbers with random phases as follows:
\begin{equation}\label{eq:RC_sigma}
\begin{split}
 w=0 &: \sigma \sim \pm 1 \ \text{with equal probability},\\
 1 \leq w < N/2 &:  \sigma_w = e^{j \theta_w}, \ \text{where $\theta_w \sim$ Uniform($[0,2 \pi]$)} \\
 w=N/2 &: \sigma_{N/2+1} \sim \pm 1 \ \text{with equal probability} \\
 N/2+1 \leq w \leq N-1 &: \sigma_w = \sigma^*_{N-w}, \text{the conjugate of $\sigma_{N-w}$}.
\end{split}
\end{equation}
From (\ref{eq:RC_sigma}) one can see that the action of $\mathbf{H}$ on a signal $\mathbf{x}$ can be broken down into a DFT followed by a randomization of the phases with symmetric constraints, followed by an inverse DFT. Fourier optics imaging architecture implementing random convolution followed by randomly pre-modulated summation (RPMS) is shown in Fig. \ref{fig:RC-diagram}. Alternatively, the random sampling process can also be substituted with randomly pre-modulated summation, which means to break them into blocks of size $N/M$, and summarize each block with a single number. This action will influence the bound of sufficient recovery measurements with a factor of $\log N$.

Random convolution is significant since it is deemed as an efficient data acquisition strategy that can recover noiseless $N$-length signals in any fixed representation from $\mathcal{O} (K \log N)$ measurements, which is relatively small for structured CS matrices. The randomness exists in both sampling process and entries generation, making RC universal (or uniform) towards the choice of signal representation. It is specially important for signals sparse in unknown bases.

\subsection{Structurally Random Matrices}

\begin{figure*}[tp]
\mbox{}\\
   \centering
   \includegraphics[width=10cm]{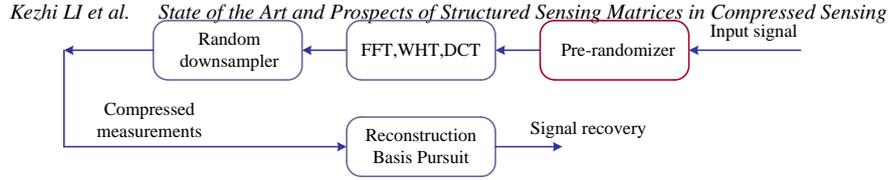} 
  \caption{Block diagram for sampling scheme of SRM \cite{Do-fast-and-efficient}.}\label{fig:SRM-diagram}
\end{figure*}

Structurally Random Matrix (SRM) is a novel framework of fast and efficient CS introduced by Do et. al. \cite{Do-fast-and-efficient,Do-ICASSP}. In the SRM, the sensing signal is prerandomized by scrambling its sample locations for flipping its sample signs and then fast-transforming the randomized samples. The sensing measurements are obtained by subsampling the resulting transform coefficients finally. The sampling algorithm contains 3 steps. The diagram is illustrated in Fig. \ref{fig:SRM-diagram}.

As shown in Figure \ref{fig:SRM-diagram}, the sampling procedure is (i) pre-randomizing a signal; (ii) applying some fast transform to the randomized signal; (iii) randomly subsampling the transform coefficients to get compressed measurements. If decomposing the algorithm mathematically as a product of 3 matrices, then the SRM can be represented as \cite{Do-fast-and-efficient}
 \begin{equation}
 \mathbf{A} = \mathbf{DFR},
 \end{equation}
where
\begin{itemize}
\item $\mathbf{R}$, the randomizer, is a random permutation matrix (denoted as the global randomizer) or a random diagonal matrix of Bernoulli i.i.d entries (denoted as the local randomizer)

\item $\mathbf{F}$ is some computable transform such as the FFT, the DCT, the WHT, ect

\item $\mathbf{D}$, the random downsampler, is a matrix composed of nonzero rows of a random diagonal matrix whose diagonal entries $D_{ii}$ are i.i.d. binary random variables with $P(D_{ii} = 1) = M/N$, where $M$ is the number of measurements.
\end{itemize}

The reconstruction algorithm can be any $l_1$ norm minimization or greedy pursuit algorithm. SRMs are highly relevant for large-scale, real-time compressed sensing applications as they have fast computation
and support block-based processing. Meanwhile, SRMs have theoretical sensing performance of $\mathcal{O}(K \log N)$ measurements for exact recovery, which is comparable to that of completely random sensing matrices. In the construction of sensing operator, SRMs use the random downsampler, fast transform and random diagonal matrix, like random convolution. SRMs provide the properties of universality and hardware implementation friendliness for reconstructing sparse signals.

\subsection{Structured matrices using sequences}
 In the most previous work, random sequences have been exploited to generate sensing matrices. \cite{Haupt-Toeplitz, Rauhut-Restricted} use Bernoulli random sequence. An alternative way
is to obtain matrices from diagonal unimodular sequences $\mathbf{\sigma}$ with
random phases \cite{random-convolution}, i.e.,
$\sigma_k=e^{j\theta_k}$, where $\theta_k$ is a random variable that
is uniformly distributed in $[0, 2\pi)$. In \cite{PRSP,Fan-TheSynthesis}, $\sigma$ can be perfect or nearly perfect sequences.

Different from random sequences, recently many researchers adopt deterministic sequences to construct sensing matrices.
 These sequences are generated delicately and many of them have been widely implemented in communication and coding theory. Because the sequences have determined the formulation, the sensing matrices based on sequences often have less randomness, and many of them are even deterministic \cite{Applebaum-Chirp, Gan-GolaymeetsHada, Kezhi-Convolutional-SP, Large-class-STRIP,Kezhi-NovelRadar}. Here we only introduce one of them named convolutional matrices using deterministic filter \cite{Kezhi-Convolutional-SP} as an example to have a look how to construct sensing matrices employing sequences.


The sampling operator ${{\bf \Phi}}$ can be represented as a partial
circulant matrix with the following form \cite{Kezhi-Convolutional-SP}
\begin{equation}\label{eq:A}
{{\bf \Phi}}= \frac{1}{\sqrt{{M}}} \mathbf{R}_{\Omega} \mathbf{A},
\end{equation}
where $\mathbf{A}$ is a circulant matrix that can be expressed as
\begin{equation}
\mathbf{A}= { \left[ \begin{array}{ccccc}
a_0 & a_{N-1} & \cdots & a_{1}  \\
a_{1} & a_0 & \cdots & a_{2} \\
\vdots & \vdots & \ddots & \vdots \\
a_{N-1} & a_{N-2} & \cdots & a_0
 \end{array} \right]}.
\end{equation}
For ${{\bf \Phi}}$ given in (\ref{eq:A}), the measurement process can be realized by circularly
convolving ${\mathbf{x}}$ with a filter $\mathbf{a}=\begin{bmatrix}a_0 & a_1 & \cdots & a_{N-1}
\end{bmatrix}^T$ and then downsample the output at locations indexed
by $\Omega$. As known the circulant matrix $\mathbf{A}$ can be diagonalized using FFT. This property enables the convolutional matrix with
fast computation. It is easy to see that the filter
vector $\mathbf{a}$ (i.e., the first column of $\mathbf{A}$) can be
obtained by taking the inverse FFT
of sequence ${\mathbf{\sigma}}=\begin{bmatrix}\sigma_0 & \sigma_1 &
\cdots & \sigma_{N-1}\end{bmatrix}^T$, i.e.,
\begin{equation}\label{eq:Ifft-sigma}
\mathbf{a}=\frac{1}{\sqrt{N}}{\mathbf{F}}^{*}\mathbf{\sigma}.
\end{equation}
$\bf{\sigma}$ may adopt various unimodular sequences. The coherence bounds for different sequences are given in Table \ref{tab:muA}.
For real sensing matrices $\mathbf{A}$, the diagonal
sequence needs to be conjugate symmetric, shown as extended sequences in Table \ref{tab:muA}.

\begin{table*}[t]
   \centering
\begin{tabular}{|c|c|c|c|}
  \hline
  & $\sigma$ & $N$ & $\mu(\mathbf{A})$\\\hline \hline
  \multirow{6}{*}{Complex matrices} & FZC      & Arbitrary & 1 \\\cline{2-4}
  & $m$-sequence & $2^k-1$, $k\in \mathbb{N}$ &   $\sqrt{1+\frac{1}{N}}$\\\cline{2-4}
  & \multirow{2}[4]{*}{Legendre sequence} & $N\equiv 3$ (mod 4) and $N$ prime & $\sqrt{1+\frac{1}{N}}$\\\cline{3-4}
  &  & $N\equiv 1$ (mod 4) and $N$ prime & $1+\frac{1}{\sqrt{N}}$\\\cline{2-4}
  & Golay sequence & $2^{\kappa_1}10^{\kappa_2}26^{\kappa_3},{\kappa_1},{\kappa_2},{\kappa_3} \in  \mathbb{N}$ & $\sqrt{2}$
  \\\hline
  \multirow{5}{*}{Real matrices} & \multirow{2}[4]{*}{Extended FZC} & Even $N$ & $4+ \frac{4}{\sqrt{N}}$\\\cline{3-4}
  &  & Odd $N$ & $2.69+ \frac{8.15}{\sqrt{N}}$ \\\cline{2-4}
  {}& \multirow{2}[4]{*} {Extended Golay} & Even $N$, $N=2^{\kappa_1}10^{\kappa_2}26^{\kappa_3},{\kappa_1},{\kappa_2},{\kappa_3} \in  \mathbb{N}$ & $2 \left( 1+{\frac{1}{\sqrt{N}}} \right)$\\\cline{3-4}
  &  & Odd $N$, $N=2^{\kappa_1}10^{\kappa_2}26^{\kappa_3}-1,{\kappa_1},{\kappa_2},{\kappa_3} \in  \mathbb{N}$ & $1+{\frac{2}{\sqrt{N}} }$ \\\hline
   \end{tabular}     
  \centering
\caption{Coherence parameter $\mu(\mathbf{A})$ for different diagonal sequences $\sigma$}\label{tab:muA}
\end{table*}

Using the uniform and non-uniform theorems, the coherence bounds reveal that $M \geq \mathcal{O}(\delta^{-2} \mu K \log^4 N)$ measurements are enough for uniform recovery and $M \geq \mathcal{O}( \delta^{-2} \mu K \log N)$ for non-uniform recovery, where $\delta$ denotes the restricted isometry constant. These convolutional matrices are not universal, while they show the effectiveness for signals sparse in both the time and frequency domain. When $\sigma$ is the Frank-Zadoff-Chu (FZC) sequence, the corresponding sensing matrices are also capable for recovering signals sparse in the DCT domain. The number of measurements in regard to different sequences can be calculated easily from Table $1$ and Theorem $1,2$.

\subsection{Other Sensing Matrices}

Many other sensing matrices were developed in recent years. To accelerate the computational speed for large data, block structures were introduced for Gaussian matrix \cite{Gan-Digital}, Toeplitz matrix \cite{Sebert-Toeplitz}, Hadamard matrix \cite{LuGan-FastCI} and SRM \cite{Do-fast-and-efficient} etc. The block structure means the sensing matrices have the following form with structured matrices as blocks $\mathbf{A}_i, i=1, \cdots, l$.
 \begin{equation}
 \mathbf{A} =  { \left[ \begin{array}{cccc}
\mathbf{A}_1 &  &  &   \\
 & \mathbf{A}_2 &  &  \\
 &  & \ddots &  \\
 &  &  & \mathbf{A}_l
 \end{array} \right]},
 \end{equation}

Block-based sensing has more advantageous for realtime applications since the encoder does not need to send the sampled data until the whole signal is measured. Besides structure sensing matrices, the sensing matrices can even be deterministic. Various deterministic matrices have been introduced in \cite{Deterministic-devore,VS-Deterministic, Iwen-deterministic, Large-class-STRIP,Applebaum-Chirp, Monajemi-Deter-PNAS}. Comparing to structured sensing matrices, deterministic sensing matrices has fixed forms and there is no randomness in the construction. Specifically, second order Reed-Muller codes are used in \cite{Howard-AFastRec,Large-class-STRIP} and dual of extended binary BCH codes are exploited in \cite{Ailon-FastDim, Large-class-STRIP}. Many other sequences are also employed in the deterministic matrix design, such as discrete chirp sequences \cite{Applebaum-Chirp, Kezhi-Convolutional-SP, Large-class-STRIP}, Kerdock and Delsarte-Goethals codes \cite{Calderbank-Compu}, Sidelnikov sequences \cite{Yu-Deter-CS-mul} and Alltop sequences \cite{Alltop-Complex,Large-class-STRIP,Strohmer-Grassmanian} etc. Deterministic sensing matrices have fixed constructions, and thus normally they can not guarantee to recover all signals with high probability. They are able to recover most signals but an exponential fraction with high probability. Some papers focus on the problems of designing sensing matrices that lead to good (expected-case) mean squared error (MSE) performance rather than the worst case \cite{Chen-ProjectionDes}. For more information regarding sensing matrices the readers may refer to references \cite{Duarte-Structured, Toeplitz_nonuniform} and CS website \cite{CS-website}.

\subsection{Relations Between Structured Sensing Matrices}

 The sensing matrices introduced in this section are not developed independently. They are associated with each other. Subsampled Fourier and Hadamard matrices were firstly proved as the qualified structured sensing matrices. They belong to the subsampled incoherent bases group of matrices. Random Toeplitz matrices are very famous and significant to many applications, such as channel estimation \cite{Haupt-Toeplitz} and system identification \cite{VS-Deterministic}. The randomness exists in each row while between rows they have strong structure. In real time signal processing the modulation idea has been widely implemented, which is also used in random demodulator. The celebrated random convolution is actually a specially modulation of signals in the Fourier domain. Moreover, SRM are a group of structured matrices generated from an approach based on random convolution but with Bernoulli diagonal phase modulation for signals in more flexible domains. Finally, structured and deterministic sensing matrices using sequences are analyzed as a new sub-area in sensing matrix design.

Practically people may utilize different structured sensing matrices according to the sensing models and hardware constraints. For instance, if one needs structured matrices to model the 1-dimensional convolution in sensing processing, random Toeplitz or Circulant matrices are employed due to the natural of the convolution calculation. In addition, in the same model if the objective signal is sparse in the Fourier domain and the phases of the modulated signal can be symmetrical, random convolution are suitable to solve this problem accelerated by fast algorithms. When the phases can only be modulated as $\pm 1$ and should be determined in advance for hardware reasons, convolutional CS matrices using deterministic sequences are recommended with the price of more number of measurements $M$ comparing with that of random convolution. If fixing the entire sensing scheme, the sensing matrices will be deterministic and there is no randomness in construction. In this case deterministic matrices constructed from coding theory are the only candidates, and usually they are with strict size constraints. In general, there is a tradeoff between randomness and number of measurements $M$. Less randomness facilitates the sensing scheme, however it often leads to more measurements and consequently longer sensing time.

\section{Applications of Structured Sensing Matrices}

Essentially CS theory can be recognized as a data processing technique that recovers sparse data from under-determined equations. The advantage of CS is to process sparse signals that can not be processed appropriately before, or obtain the compressed data using proper physical instruments directly. Fortunately most of the signals in the real world that people are interest in belong to sparse signals or can be approximated in certain domain. So from its emergence CS has been implemented in numerous applications including communications, machine learning, imaging, geophysical data analysis, radar, remote sensing, data streaming, quantum state tomography, and so on. For instance the matrices mentioned previously, the Toeplitz matrices are quite suitable for communication channel estimation \cite{Haupt-Toeplitz}; random demodulators are designed for sampling of sparse wideband analog signals\cite{Tropp-BeyondNyquist, Mishali-From}; random convolution matrices can be exploited in radar imaging \cite{random-convolution}; also, the validity of SRMs has been verified in image processing\cite{Do-fast-and-efficient,Do-ICASSP}; convolutional matrices using sequences have widely applications in communication and signal processing \cite{Kezhi-ICASSP2011,Kezhi-Convolutional-SP,Kezhi-NovelRadar,Kezhi-Wyner-DCS}. Apart from these works, here we simply introduce two celebrated applications of CS, in medical imaging and single pixel camera.

\begin{figure*}[ht]
   \centering
   \begin{minipage}[t]{0.49\linewidth}
   \centering
  \includegraphics[width=6cm]{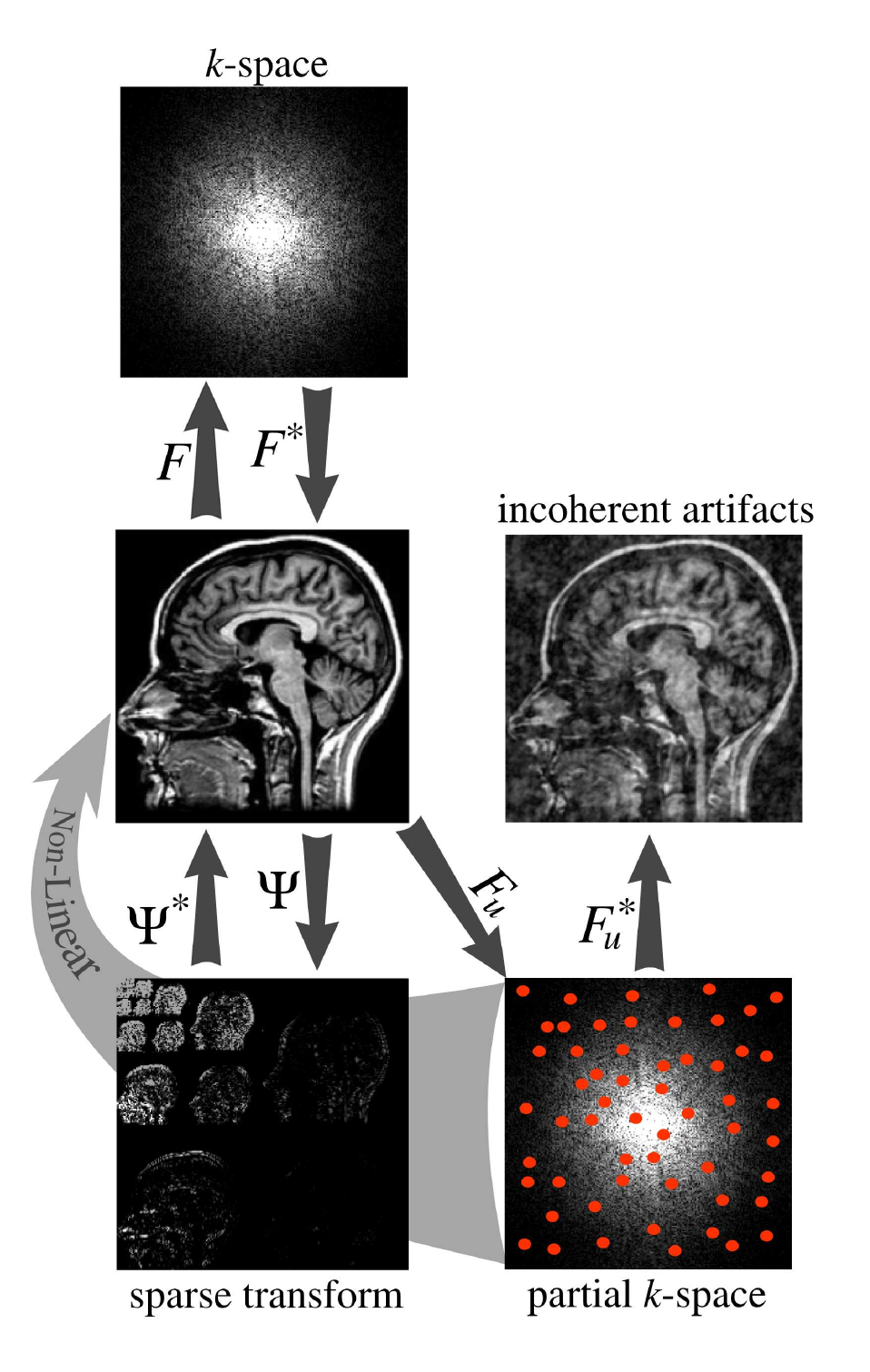} 
\centerline{(a)} 
   \end{minipage}
   \begin{minipage}[t]{0.49\linewidth}
   \centering
   \includegraphics[width=3.8cm]{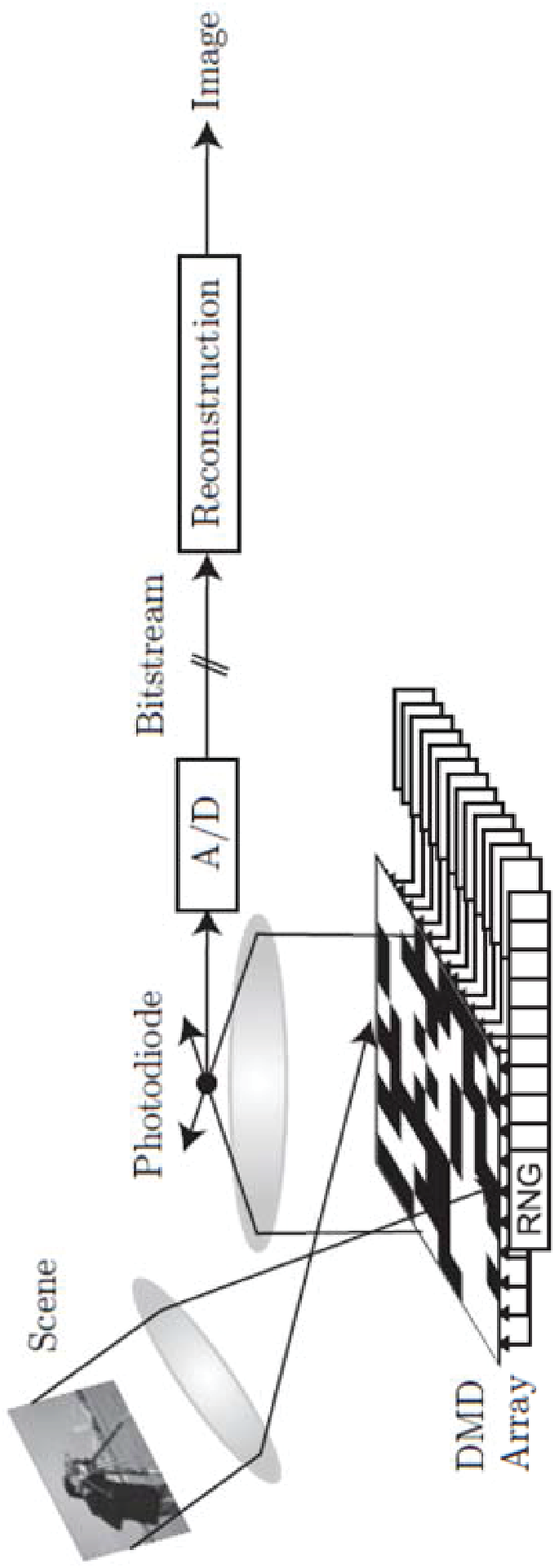} 
\centerline{(b)} 
   \end{minipage}

  \caption[(a) Illustration of the domains and operators as the requirements
of CS: sparsity in the transform domain, incoherence of the undersampling artifacts and the need
for nonlinear reconstruction that enforces sparsity. (b) Diagram of the single pixel camera. The image $\mathbf{x}$ is reflected off a digital micro-mirror device (DMD) array whose mirror orientations are modulated in the pseudorandom pattern supplied by the random number generator (RNG).]{(a) Illustration of the domains and operators used in \cite{Lustig-SparseMRI} as the requirements
of CS: sparsity in the transform domain, incoherence of the undersampling artifacts and the need
for nonlinear reconstruction that enforces sparsity. (b) Diagram of the single pixel camera. The image $\mathbf{x}$ is reflected off a digital micro-mirror device (DMD) array whose mirror orientations are modulated in the pseudorandom pattern supplied by the random number generator (RNG) \cite{Singlepixel_jnl}.}\label{fig:Appli1}
\end{figure*}

A promising application for compressed sensing is in reducing the sampling rate
in magnetic resonance imaging (MRI) \cite{MRI-CS,Gorodnitsky-Neu, Lustig-SparseMRI}. The main motivation of CS MRI is that,
 MRI scanners sequentially sample the human's body in the 2-D continuous Fourier domain, and sensed coefficients satisfy the sparse property which is also the prerequisite of the theory of CS. Moreover, MRI is very time costly. In order to obtain a clearer image, one often needs a long time to collect the data. However, the speed of data collection is limited by physical and physiological constraints. Applying the CS technique may accelerate the scanning process with the same accuracy due to fewer CS measurements being required. The schematic diagram of MRI using CS is shown Fig. \ref{fig:Appli1} (a).

Another category of the application involves the design of new acquisition hardware that is able to acquire projections of a signal against a class of vectors. In this case, the sensing process is accomplished by physical optical instruments, and the research normally focuses on the problem of how to design sensing matrices whose entries belong to some patterns/bases that can be easily implemented on the hardware. One example is the framework of recovering an image based on optical modulators, known as the single pixel camera shown in Fig. \ref{fig:Appli1} (b) \cite{Singlepixel_jnl}. The digital micromirror device (DMD) is a reflective spatial light modulator that selectively redirects parts of the light
beam \cite{Sampsell-AnOver}. The DMD is comprised of an array of bacterium-sized,
electrostatically actuated micro-mirrors, and each mirror rotates
about a hinge and can swing between two stages $+10^\text{o}$ or $-10^\text{o}$. The state of each mirror depends on the bit loaded in the corresponding position of the programmable sensing matrix, and many structured sensing matrices may be implemented in this scenario. People have tested that the system works well when matrix entries are drawn randomly from a fast transform such as a Walsh Hadamard transform \cite{Takhar-computational}. Many advanced imaging hardware architectures based on the single pixel camera model have been developed after these techniques mature, e.g. in terahertz imaging \cite{Chan-Terahertz, Shen-CTHzImaging}. With regard to real applications, actually it is not trivial to decide which strategy or structured sensing matrices we shall use. Because different matrices have their own features and performances, we have to investigate the practical scenarios and make the tradeoff between number of measurements $M$, universality or not, hardware constraints, computation and so on.

\section{Prospects And Future Works}

As the key research area of the encoding part of the compressed sensing theory, the research of structured sensing matrices is really important and has attracted more and more attention in recent several years. Although in literatures people have proposed many structured sensing matrices, matrices with special structures are deadly needed with regard to special settings or hardware requirements.

Generally speaking, the future development of sensing matrices will focus on two aspects. The first one is to use less randomness and less memory storage. For instance, comparing with full random matrices, more sparse sensing matrices with certain structure have and will be exploited to reduce the calculation in CS \cite{Gilbert-SparseRe}. The structure of a network also can be embodied in a matrix revealed by a one-to-one correspondence with an expander graph \cite{Xu-EfficientCom}. The second aspect is to design sensing matrices satisfying certain structure in reality. This will be the main motivation for developing more structured sensing matrices. Take several examples to illustrate it. In the communication system the convolution process is equivalent to a Toeplitz matrix consisting of the transmitting signals multiplying the system impulse response function. That's the reason why Toeplitz CS matrices could be utilized in sparse channel estimation \cite{Haupt-Toeplitz}. In \cite{Mishali-From} the authors proposed a practical sampling system called modulated wideband converter (MWC) by adopting periodic waveforms, a low-pass filter and a low rate sampler. They proved that perfect recovery of multi-bandlimited signals from the proposed samples can be achieved under certain necessary and sufficient conditions. In mathematics, the sampling process can be reformed as a structured sensing matrices $\mathbf{y} = \mathbf{S \bar{F} D}$, where $\mathbf{S \bar{F} D}$ represent the sign pattern matrix, reorder Fourier matrix and diagonal matrix, respectively. This matrix structure comes from the hardware design, and it performs well in practice \cite{Mishali-From}. In addition, the structured sensing matrices were also implemented in statistical physics, such as the seeding matrix with coupling block diagonal structure. This work was proposed in \cite{Krzakala-Statistical-Phy} for a framework named seeded compressed sensing. Krzakala et. al. proved that in their model the experimental recovery results approached the theoretical limit for large systems. To sum up, people will continue to work on pursuing various structured sensing matrices with less randomness/measurements, better performances and hardware friendly property cooperating knowledge from other fields such as coding theory, communication, random matrix theory etc., subject to specific requirements based on real settings.

\section{Conclusion}
After explaining the fundamental knowledge of sensing matrices, RIP and coherence, we reviewed couples of existing structured sensing matrices, including subsampled incoherent bases, random Toeplitz matrices, random demodulator matrices, random convolution matrices, structurally random matrices and other structured sensing matrices. For each of them, we concentrated on the structure of the matrix, the measurement bounds and its scope of application. Though it is difficult to cover all of the developments in structured sensing matrices area, here we aim to explain the main idea and demonstrate a few well-known examples that are representatives of a wider class of the CS problem. 

\section*{Acknowledgment}

This work was partially supported by the Swedish Research Council, the Linnaeus Center ACCESS at KTH, the European Research Council under the advanced grant LEARN, contract 267381, and the China National Key Basic Research Program under Grant No. 2011CBA00200.

\bibliographystyle{IEEEtran}

\end{document}